\documentclass[11pt]{article}

\usepackage{epsfig}
\usepackage{subfigure}
\usepackage{times}
\usepackage{amsfonts}
\usepackage{amssymb, amsmath}
\usepackage{epsfig}
\usepackage{setspace}

\usepackage{times}
\newcommand{\ignore}[1] {}
\newcommand{\sgemm} {{\tt spgemm}}
\newcommand{\spmv} {{\tt spmv}}
\newcommand{\bilat} {{\bf Bilat}}
\newcommand{\conv} {{\bf Conv}}
\newcommand{\hist} {{\bf hist}}
\newcommand{\sort} {{\bf sort}}
\newcommand{\LR} {{\bf LR}}
\newcommand{\CC} {{\bf CC}}
\newcommand{\RC} {{\bf RC}}
\newcommand{\MC} {{\bf MC}}
\newcommand{\Bundle} {{\bf Bundle}}
\newcommand{\Dither} {{\bf Dither}}
\newcommand{\LBM} {{\bf LBM}}

\title{CPU and/or GPU: Revisiting the GPU Vs. CPU Myth}
\author{Kishore Kothapalli, Dip Sankar Banerjee, P. J. Narayanan,
Surinder Sood,\\ Aman Kumar Bahl, Shashank Sharma, Shrenik Lad, Krishna
Kumar Singh, Kiran Matam,\\ Sivaramakrishna Bharadwaj, Rohit Nigam,
Parikshit Sakurikar, Aditya Deshpande,\\ Ishan Misra, Siddharth Choudhary, Shubham Gupta\\
International Institute of Information Technology, Hyderabad \\
Gachibowli, Hyderabad 500 032, India. 
}
\date{}

\begin{document}

\maketitle

\begin{abstract}
Parallel computing using accelerators has gained widespread research
attention in the past few years. In particular, using GPUs for general
purpose computing has brought forth several success stories with respect
to time taken, cost, power, and other metrics. However, accelerator
based computing has significantly relegated the role of CPUs in
computation. As CPUs evolve and also offer matching computational
resources, it is important to also include CPUs in the computation. We
call this the {\em hybrid computing} model. Indeed, most computer
systems of the present age offer a degree of heterogeneity and therefore
such a model is quite natural.

We reevaluate the claim of a recent paper by Lee et al.(ISCA 2010). We
argue that the right question arising out of Lee et al. (ISCA 2010)
should be how to use a CPU+GPU platform efficiently, instead of whether
one should use a CPU or a GPU exclusively. To this end, we 
experiment with a set of 13 diverse workloads
ranging from databases, image processing, sparse matrix kernels, and graphs.
We experiment with two different hybrid platforms: one consisting of a
6-core Intel i7-980X CPU and an NVidia Tesla T10 GPU, and another consisting
of an Intel E7400 dual core CPU with an NVidia GT520 GPU. On both these
platforms, we show that hybrid solutions offer good advantage over CPU or GPU alone
solutions.  On both these platforms, we also show that our solutions are
90\% resource efficient on average. 

Our work therefore suggests that hybrid computing can offer tremendous
advantages at not only research-scale platforms but also the more realistic
scale  systems with significant performance gains and resource efficiency to the 
large scale user community. 

\end{abstract}

\section{Introduction}
\label{sec:intro}
Parallel computing using accelerator based platforms has  
gained widespread research attention in recent years. 
Accelerator based general
purpose computing, however, relegated the role of CPUs to second-class
citizens where a CPU sends data and program to an accelerator and
gets the results of the computation from the accelerator. As CPUs
evolve and narrow the performance gap with accelerators on
several challenge problems, it is imperative that parity is restored
by also bringing CPUs in the computation process. 

\begin{figure}
\centering
\includegraphics[scale=0.25]{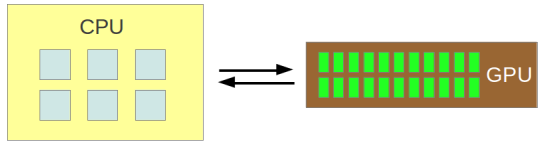}
\caption{A tightly coupled hybrid platform.}
\label{fig:cpugpu}
\end{figure}

Hybrid computing seeks to
simultaneously use all the computational resources on a given
tightly coupled platform. We envisage a multicore CPU plus accelerators 
such as GPUs, see also Figure \ref{fig:cpugpu}, as one such realization. 
The CPU in Figure \ref{fig:cpugpu} on the left with six cores is
connected to a many-core GPU on the right. In our work, we use two
variants of the model shown in Figure \ref{fig:cpugpu} by choosing 
different CPUs and GPUs.

\begin{figure}[htp]
  \begin{center}
    \includegraphics[scale=0.15]{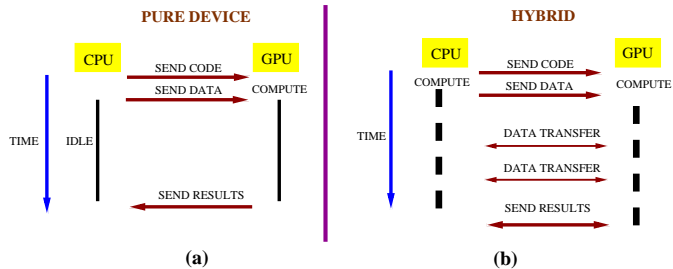}
  \end{center}
  \caption{A view of hybrid multicore computing. Figure (a) shows the
  conventional accelerator based computing where the CPU typically stays
  idle. Figure (b) shows the hybrid computing model with computation
 overlapping between the CPU and the GPU.}
  \label{fig:hybrid}
\end{figure}

The case for hybrid computing on such a platform can be made naturally. 
Computers come with a CPU, at least a dual-core at present, and is expected
to contain tens of cores in the near future. Graphics processing
units are traditionally used to process graphics operations
and most computers come equipped with a graphics card that presently
has several GFLOPS of computing power. Moreover, commodity production
of GPUs has significantly lowered their prices. Hence, an application using both a
multicore CPU and an accelerator such as a GPU can benefit from
faster processing speed, better power and resource utilization, and
the like.  Figure \ref{fig:hybrid} illustrates the benefits of
such a hybrid computing
model. As can be noticed in Figure \ref{fig:hybrid}(b), hybrid computing
calls for complete utilization of the available computing resources.

Further, it is believed that GPUs are not well suited for computations
that offer little SIMD parallelism, and have highly irregular memory
access patterns. For various reasons, CPUs do not suffer greatly on such
computations. Thus, hybrid computing opens the possibility of novel
solution designs that take the heterogeneity into account. Hence, we
posit that hybrid computing has the scope to bring the benefits of
high performance computing to also desktop and commodity users.

We distinguish between our model of hybrid computing with other existing
models as follows. We consider computational resources that are tightly
coupled. Supercomputers such as the Tianhe-1A that
use a combination of CPUs and GPUs do not fall in our category. The
issues at that scale would overlap with some of the issues that arise in
hybrid computing, but have other unique issues such as interconnection
network and its cross-section bandwidth, latency,  and the like. 

\ignore {
Hybrid computing brings its own set of challenges
to algorithm design, algorithm engineering, analysis, optimization,
and other system related challenges. Some of the challenging
issues that are relevant to hybrid computing are:

\begin{itemize}
\item Designing hybrid algorithms that are efficient, scalable,
and resource-friendly,
\item Implement, test correctness, reason about, and debug
hybrid solutions
\item Analyse hybrid algorithms, and 
\item Characterization of problems that allow for efficient hybrid
algorithms? 
\end{itemize}
}

Hybrid computing solutions on platforms using a combination of CPUs and GPUs
are being studied for various problem classes such as dense linear algebra
kernels \cite{BDT08, dongarra2009}, maximum flows in networks \cite{hong09},
list ranking \cite{WJ10} and the like. 
Most of these works however have a few limitations. In some cases,
for example \cite{hong09,WJ10, SMI12}, computation is done only
either on the CPU or the GPU at any given time. Such a scenario keeps one
of the computing resource idle and hence is in general wasteful in
resources. Secondly, we explore a diverse set of workloads so as to
highlight the advantages and limitations of hybrid multicore computing.
In a departure from most earlier works, we study hybrid computing also on
low end CPU and GPU models and see the viability of hybrid computing on
such platforms.

The aim of this work is to explore the applicability of hybrid
multicore computing to a class of applications. To this end, we
select a collection of 13 different workloads ranging from databases
to graphs, and image analysis, and present hybrid algorithmic
solutions for these workloads. Some of the results that we show in this
paper are reported recently \cite{lspp12,hipc11,hipc11pjn,bundle,sgemm}, or
under submission \cite{newsort}.  We 
identify two different algorithm engineering approaches that we
have used across our 13 workloads. Further, these approaches can be 
used to classify most of the recent works on hybrid computing
including \cite{dongarra2009, AICCSA11}. The two
approaches are described briefly in the following:

\begin{itemize}
\item {\bf Work Sharing:} In a work sharing approach where the problem is 
decomposed into one or more parts, with each part running on a different 
machine in the hybrid computing platform. In this approach, the work shares
have to be chosen appropriately so as to balance the load on the CPU and the
GPU. In this approach, the actual  algorithm used on the  different units
could be different, and in some  cases, may also be different from the best
possible algorithm on each  device respectively. For example, see our hybrid
algorithm for graph connected components and sparse matrix-vector
multiplication described in Section \ref{sec:impl}.
\ignore{(For instance, consider the \spmv$\;$ kernel which involves
multiplying a sparse matrix with a vector to produce another vector.
The best hybrid algorithm may actually assign rows that the CPU is
good at working with to the CPU and vice-versa on the GPU. This may
be different from the best possible pure CPU/GPU algorithm which has
to compute on all rows.)}

\item {\bf Task Parallel:}  A task parallel approach involves viewing the 
computation as a collection of inter-dependent tasks, and tasks are 
mapped on to the available units. In this setting, one has to think of
arriving at the best possible mapping that minimizes the overall span and 
also minimizes the idle time of machines. As can be seen, the time taken 
by a hybrid solution using task parallelism is the time corresponding to 
the longest path in the task graph with nodes and edges labelled by the time
taken to complete the task and communication time respectively.

A slight modification to the task parallelism approach is that of pipelined
parallelism.
A pipelined parallelism approach views the computation as a 
clever combination of the units to set up a pipeline that has
different functionalities at each stage of the pipeline. 
\end{itemize}

\subsection{Our Contributions}

Some of the specific contributions of our study are as follows.

\begin{itemize}

\item We develop new hybrid solutions for seven of the 13 workloads studied
in this paper: sorting,
histogram, \spmv, bilateral filtering, convolution, ray casting, and
the Lattice Boltzman Method. 

\item We experiment with a hybrid system consisting of a six core Intel
i7 980X (Westmere) CPU and an NVidia Tesla T10 GPU. On this hybrid platform, called
{\em Hybrid-High}, we show
that hybrid computing has an average of 29\% performance improvement
over the best possible GPU-alone implementation. Our solutions also
exhibit 90\% resource efficiency on average.

\item To promote the idea that hybrid computing has benefits on 
more widely used platforms, we also experiment with a hybrid
platform that has an Intel Core 2 Duo E7400 (Allendale) with an NVidia GT520 GPU. 
We feel that such configurations still have the potential to make
supercomputing affordable and accessible to everyone. 
On such a platform, called {\em Hybrid-Low}, we show that hybrid computing
results in an average of
37\% performance improvement compared to the best possible GPU-alone
implementation.  Our solutions also exhibit 90\% resource efficiency on average.

\item We analyze the above results and offer insights on the limits and
applicability of hybrid computing in the present architectural space. 
\end{itemize}

The title of our paper is
motivated by a recent related paper \cite{isca10} that argued about the relative
strengths of GPUs and multicore CPUs on a set of throughput oriented
workloads.  We establish through this paper that accelerator based
computing should leverage the combined strengths of all devices in a
computing platform via hybrid computing. A majority of our workload overlap
with the workloads considered in \cite{isca10}.

\subsection{Organization of the Paper}
The rest of the paper is organized as follows. Section \ref{sec:prelim}
discusses the architectures of the GPUs and the CPUs used in our paper. 
Section \ref{sec:workloads} describes the workloads that we study in this
paper. Section \ref{sec:impl} outlines some of the important implementation
perspectives of our hybrid algorithms. Section \ref{sec:results} discusses
the outcomes of our hybrid solutions and analyzes the solutions. The work
is put in the right context by discussing related prior work in Section
\ref{sec:related}. The paper
ends with concluding remarks in Section \ref{sec:concl}.

\section{Hybrid Computing Platforms}
\label{sec:prelim}
In this section, we briefly describe the two hybrid CPU+GPU computing
platforms that we use in our study.
\ignore {
\subsection{A Brief Overview of NVidia GPUs}
\label{sec:gpu}
Nvidia's unified architecture  for its current line 
of GPUs supports both graphics and general computing. In general
purpose computing, 
the GPU is viewed as a massively multi-threaded architecture containing
hundreds of processing elements ({\em cores}). Each core comes with a
four stage pipeline. Eight cores, also known as {\em Symmetric Processors} (SPs)
are grouped in an SIMD fashion into a
{\em Symmetric Multiprocessor} (SM), so that each core in an SM executes
the same instruction.  Each core can store a
number of thread contexts. Data fetch latencies are tolerated by
switching between threads. Nvidia features a zero-overhead scheduling system
by quick switching of thread contexts in the hardware.

The GPU also has various memory types at each level. A set of
32-bit registers is evenly divided among the threads in each SM.
16 Kilobyte of {\em shared memory} per SM acts as a user-managed
cache and is available for all the threads in a Block. The GTX
280 is equipped with 1 GB of off-chip {\em global memory} which can be
accessed by all the threads in the grid, but may incur
hundreds of cycles of latency for each fetch/store. 
}

\subsection{The Hybrid-High Platform}
One of the hybrid computing platform we use in this paper, labeled {\em
Hybrid-High}, is a combination of a six core Intel i7 980X CPU with an
NVidia Tesla T10 GPU. 

The Intel i7 980X CPU that we use in our experiments is a six
core machine with each core running at 3.4 GHz and with a thermal
design power of 130 W. The six physical cores are hyper-threaded so
that together they can run 12 threads. Other features of the i7 980X include
a 32 KB instruction + 32 KB data L1 cache per core, a 256 KB L2 cache
per core, and a large shared 12 MB L3 cache shared by all 6 cores. The
memory bandwidth is up to 1066 MHz. 

The GPU is a massively multi-threaded processor
containing hundreds of processing elements or cores,
called the Scalar Processors (SPs). The Tesla C1060
is add-on card based on the Tesla T10 GPU\cite{tesla_spec} having 
SPs arranged in groups of eight. It has 30 such SMs,
which makes for a total of 240 processing cores. Each
of the cores are clocked at 1.3 Ghz. These eight SP
execute in a Single Instruction Multiple Thread (SIMT)
fashion. Hence, all the SPs in an SM execute the same
instruction at the same time.

The CUDA API allows a user to create a large number
of threads to execute code on the GPU. Threads are also
hierarchically grouped into blocks and grids. Blocks
are serially assigned for execution on each SM. Each of
the blocks are made of several warps which execute in each
core in a SIMD fashion. For more details, we refer the interested
reader to \cite{cuda}.

\subsection{The Hybrid-Low Platform}
We also experiment with a hybrid platform, called {\em Hybrid-Low}, 
that resembles a desktop computing environment more closely. 
The Hybrid-Low platform is a combination of an 
Intel Core 2 Duo E7400 CPU along with an NVidia GT520 GPU. 

The Intel Core 2 Duo CPU is one of the earliest multicore offerings from
Intel and was released in the year 2008. It has 2 cores with hyper-threading
and each of the cores are clocked at 2.8 GHz. The CPU consists of a 3 MB
L2 cache and the maximum power consumption is around 65 W. The CPU was
designed entirely for commodity PCs which gives a sustained performance
of about 20 GFLOPS. 

The GT520 is a stand-alone graphics processor having 48 computing cores 
and 1 GB of global memory. Each of the compute cores are clocked at 810 MHz.
The GPU on an average give a sustained performance of 77.7 GFLOPS and consume
about 29W of power. In this system both the processors are of a
comparable performance range
and hence provide a more realistic platform for experimenting the hybrid
programs.

\ignore{
\begin{table*}
\begin{center}
\begin{tabular}{|l|l|l|l|l|l|l|l|l|}
\hline
{\bf Name} & {\bf Memory B/w} & {\bf Freq.} & {\bf Peak SP} & 
{\bf Memory/RAM} & {\bf L2 cache}  &  {\bf \#Cores}\\
  & GB/s & GHz & GFlops & GB &  & \\
& & & & & &  \\
\hline
NVidia Tesla T10 GPU & 142 & 1.3 & 933 & 4  & 16KB  & 240\\
\hline
Geforce GT520  & 14.4 & 1.6 & 155 & 1 & 48KB &  48\\
NVidia GPU  & & & & & &   \\
\hline
Intel i7 980X CPU & 25.6 & 3.33 & 80 & 24 & 12MB &  6  \\
\hline
Intel Core 2 Duo  & 8.05 & 2.8 & 22.4 & 16 & 3MB &  2\\
E7400 CPU & & & & & &  \\
\hline
\end{tabular}
\end{center}
\caption{An architectural comparison of the systems used.}
\end{table*}
}

\section{Workloads}
\label{sec:workloads}
In this paper, we experiment with the following workloads. Table
\ref{table:workloads} summarizes the characteristics of the various workloads
considered.

\begin{table*}[htp!]
\begin{center}
\begin{tabular}{|c||l|l|l|l|}
\hline
{\bf Workload} &  {\bf Application Area} & {\bf Nature}
& {\bf Characteristics}  & {\bf Solution} \\
 {\bf (Short name)} & & & & {\bf Methodology}  \\
 \hline
\hline
 Sorting  &  Semi-numerical, & regular & compute bound & work sharing+ \\ 
{\bf sort}&   databases & & & task parallel  \\
\hline
 Histogram  &  image processing & Atomics, & memory bound & work
sharing \\
{\bf hist} & & & irregular &  \\
\hline
Sparse matrix-vector  & Sparse Linear Algebra & irregular & memory
bound & work sharing\\
multiplication  & & & & \\
\spmv & & & &  \\
\hline
Sparse matrix-matrix &  Sparse Linear Algebra & irregular &
device storage, & work sharing \\
multiplication  & & & & memory bound \\ 
\sgemm & & & & \\
\hline
 Ray casting  &  Image processing & irregular & compute bound & work sharing \\
\RC & & & & \\
\hline
 Bilateral filtering  &  Image processing & regular & compute bound
 & work sharing + \\
\bilat   & & & & task parallel  \\
\hline
 Convolution  &  Image processing & regular & compute bound & work sharing \\
\conv & & & & \\
\hline
 Monte-carlo  &  Physics, computational & regular, & compute bound,
 & task parallel \\
{\bf MC} &   finance & pseudorandom & & \\
  & & numbers  & & \\
\hline 
 List Ranking &   graphs and trees & irregular & memory bound & task
 parallel  \\
\LR  & & & & \\
\hline
 Connected Components &  graph algorithms & irregular & memory bound &
 work sharing  \\
\CC & & & & \\
\hline
 Lattice Boltzman Method &  Computational Fluid & irregular
 & memory bound & task parallel \\
\LBM   & Dynamics & & & \\	 
\hline
 Image Dithering  &  Image processing & irregular & causal dependencies &work sharing   \\
\Dither & & & & \\
\hline
 Bundle adjustment  &   Image processing & irregular & memory bound & task parallel\\
\Bundle & & & & \\
\hline
\end{tabular}
\caption{Various workloads considered in this paper.}
\end{center}
\label{table:workloads}
\end{table*}

{\bf Sorting: } Sorting is one of the fundamental operations in information
processing that has many applications. 
Recent results
on efficient implementations of sorting are reported in
\cite{sanders10,owens11}, to name a few. 
This workload becomes an important
case study due to the large number of applications. 
For the purposes of
this paper, we focus on comparison-based sorting techniques and leave
non-comparison based sorting technique such as radix sort out of the
scope.

{\bf Histogram: } One of the important operations in image processing is to
compute the histogram of the intensity values of the pixels. 
Computing the histogram of a
dataset in parallel typically requires the use of atomic operations. 
The nature of this workload such as use of atomic
operations and being memory bound make this a good case study for hybrid
computing.

{\bf  Sparse Matrix-Vector Multiplication (\spmv): } Efficient operations
involving sparse matrices are essential
to achieve high performance across numerical applications such as climate 
modeling, molecular
dynamics, and the like.
In most of the above applications, \spmv$\;$ computation is the main
bottleneck. 
Hence, efforts to speed up this computation on modern architectures have attracted
significant research attention in recent years \cite{vuduc07,lanczos}.
Further, the computation involved in \spmv$\;$ is highly irregular in nature 
due to the sparsity of the matrix. 

{\bf Sparse matrix-matrix multiplication (\sgemm): } Another operation that is
important with respect to sparse matrices is that of multiplying two sparse
matrices. This workload has found applications spanning several areas such
as graph algorithms \cite{kunle11}, numerical analysis including computation
fluid dynamics \cite{cfd1,cfd2,lanczos} and is also included
as one of the seven dwarfs in parallel computing in the Berkeley report
\cite{landscape}. Some of the recent works that have reported efficient
sparse matrix multiplication on modern architectures are \cite{buluc}. It
is generally accepted that the difficulties of this workload include its
irregular nature of computation, the difficulty in predicting the size of
the output and the concomitant memory management problems.

Both \spmv$\;$ and \sgemm$\;$ are important linear algebra kernels with
significant applications, and hence their choice is justified.

{\bf Ray casting: } This is a fundamental problem in image analysis and
computer graphics. Recent applications to medical image analysis are also
reported \cite{rcmed}. 
As all rays perform the computations independently, the problem is very much portable for parallel architectures. 
Tracing multiple rays in an SIMD fashion is challenging, because rays access non-contiguous
memory locations, resulting in incoherent and irregular memory accesses. 
The choice of this workload is justified because of the range of
applications of ray casting to visual computing.

\ignore {
\item FFT : Fast Fourier Transform, or FFT in short, is a fundamental image
processing primitive with applications to many fields such as digital
signal processing, engineering, astronomy, and the like. Given its
importance, the FFT workload is often used as a benchmark program in
performance analysis of parallel computers and the high performance
computing community. 
}

{\bf Bilateral filter: } Bilateral filter is an edge-preserving and
noise reducing filter used in image processing. It is a non-linear filter
in which intensity value at any pixel is equal to a weighted sum of
intensities in the neighbourhood. The
filter involves transcendental operations like computing exponentials,
which can be very computationally expensive. Hence, it is a compute bound
problem with regular memory access. 

{\bf Convolution: } Convolution is a common operation used in image
processing for effects such as blur, emboss and sharpen. Given the image
signal and the filter, the output at each pixel is equal to the weighted
sum of its neighbours. Since each pixel can be computed independently by a thread,
there is ample parallelism available. The computations increase with the
size of filter and exhibits high compute to memory ratio.

The bilateral filtering and convolution workloads are commonly used
filters in image processing applications, justifying their inclusion on
our workloads.

{\bf Monte-carlo: } Monte Carlo methods are used in several areas of science to
simulate complex processes, to validate simpler processes, and
to evaluate data. In Monte Carlo (MC) methods, a stochastic
model is constructed in which the expected value of a certain
random variable is equal to the physical quantity to be
determined. The expected value of this random variable is then
determined by the average of many independent samples representing
the random variable. This workload exhibits a regular memory access pattern 
and is compute bound, typically. 
The choice of this workload is justified by the wide body of
applications using Monte Carlo methods across varied domains such as
computational finance, physics, and engineering.

{\bf List Ranking: } The importance of list ranking to parallel computing
has been identified by Wyllie as early as 1978 in his Ph. D. thesis
\cite{W79}. The list ranking problem is to find the distance of every
node from one end of the given linked list. 
The workload is memory bound due to
the highly irregular nature of the computation involved. The workload is
chosen as a case study as list ranking is often a primitive in several
graph and tree based computations.

{\bf Connected Components: } Finding the connected components of a given
undirected graph has been a fundamental graph problem with several
applications. 
Ideas used
in parallel algorithms for connected components find immediate application
to other important graph algorithms such as minimum spanning trees and the
like. Hence, this workload is important and offers good scope as a case
study. 

{\bf Lattice Boltzman Method: } Lattice Boltzman Method (LBM) refers to a class of
applications from computational fluid dynamics (CFD) and are used in fluid
simulations. It is a numerical method that solves the Navier--Stokes equation 
via the discrete Boltzman equation. 
In this work, we study the D3Q19 lattice
model where over a three dimensional cubic lattice, each cell computes the new
function values based on its 19 neighbors \cite{lbm1}. 
The LBM operation is highly data parallel. This workload is considered
for its applications to computational fluid dynamics.

{\bf Image Dithering: }
In Floyd-Steinberg Dithering (FSD) (see \cite{hipc11pjn}), we approximate 
a higher color resolution image using a limited color palette by diffusing 
the errors of threshold operation to the neighboring pixels according to a
weighted matrix. 
The problem is thus 
inherently sequential and poses enormous challenge for a parallel 
implementation, let alone a hybrid implementation. Dithering has
various applications such as printing, display on low-end LCD or mobile
devices, visual cryptography, image compression, and the like. Further
this workload is significant because of its atypical nature amongst
image processing applications that does not offer embarrassing
parallelism.

{\bf Bundle Adjustment: }It refers to the optimal adjustment of bundles of
rays that leave 3D feature
points onto each camera center  with respect to both camera positions and point
coordinates. 
Bundle Adjustment is carried out using the Levenberg-Marquardt (LM)
algorithm \cite{b3,b4} because of its effective damping strategy to converge quickly
from a wide range of initial guesses.
Bundle adjustment often is the slowest and the most computationally resource
intensive step. It is one of the primary bottleneck in the
Structure-from-Motion pipeline, consuming about half of the total computation time.

\section{Implementation Details}
\label{sec:impl} 
In this section, we describe the implementation details of the various
workloads described in Section \ref{sec:workloads}. For the workloads
including \sort, \bilat, \conv, \spmv, \hist,
\RC, and \LBM$\;$ we developed hybrid implementations for the  purposes of
this paper. In some cases, we use implementation developed in recent existing
works that are known to be the
best possible. Workloads \LR, \CC, \Dither, \MC, \sgemm, and \MC$\;$ fall under this
category.   For more details on these implementations, we refer the reader
to the technical reports available at \cite{url}.

\subsection{Sorting}

Our implementation of sorting is a comparison based sorting
algorithm based on the techniques of sample sort reported in
\cite{sanders10}. Sample sort involves placing the elements into various
bins according to a number of splitters.
For sorting, we apply the basic principle of work partitioning
in a hierarchical fashion. We, first compute the histogram of
the data in a hybrid manner. Using the histogram results we
perform the binning process. As, the histogram provides a good
estimate of the distribution of the data, the binning process
consumes a much lesser overhead. However, on the initial iteration
of the kernel, the individual bins that are created are large in size
and cannot be directly used for sorting. 
In order to optimally sort each of the bins, we  reduce the size of
each of the bins down to a certain threshold where groups of 32
elements can be compared by a single warp. Each of these warps will in effect
implement quick sort on the 32 elements. We recursively run the binning
process to reduce the bin sizes to the chosen threshold.  The CPU on
the other hand,
will not be limited by the compute as much as the GPU. We can hence
still leave the bin
sizes of the CPU at a higher threshold than that of the GPU. 
We also notice that there is a clear
trade-off in the number of recursive calls to split the elements
into bins, and the time taken to sort the bins independently.

\subsection{Histogram}
The histogram operation in a parallel setting requires the
proper use of atomic increment operations to ensure consistent
and reliable results. We use the work sharing approach
where we divide the data set into two sets for the GPU and
the CPU. We then perform the computation of the histogram on both
the devices in an overlapped fashion. This step is followed by a simple
addition of the results from the two devices. On the GPU, using shared memory
is critical in order to reduce the global memory latency. The atomic
increment is performed by a single warp working on the data that is
obtained from the shared memory. The histogram in general is a bandwidth
bound problem and hence, proper use of the memory channels are essential.
The shared local cache (L1) in the CPU is used to improve the performance.
The resulting partial  histograms are then
added bin-by-bin to give the final histogram result over the
entire input data.

\subsection{\spmv}
In the \spmv $\;$ workload, some of the challenges faced by modern architectures
include the overhead of the auxiliary data structures, irregular memory
access patterns due to the sparsity of the matrix, load balancing, and the
like. Several recent works \cite{BG09} have therefore focussed on optimizing
the \spmv$\;$ computation on most modern architectures. 

In our hybrid implementation, we use a novel work sharing based solution 
summarized as follows. 
In \spmv $\;$, we notice that the computation involving one row is independent of
the computation involving other rows. This suggests that one should attempt a
work sharing based solution. Instead of splitting the computation according to
some threshold, we use  the following novel work sharing approach. Our
approach is guided by the fact that typically \spmv $\;$ is used over multiple
iterations. So, one can rely on preprocessing techniques that aim to improve
the performance of \spmv. 

Notice that the GPUs are good at exploiting massive data parallelism
with regular memory access patters. Therefore, we assign the
computation corresponding to the dense rows to the GPU and the computation
corresponding to the sparse rows to the CPU. The exact definition of
sparsity is estimated via experimentation. In this direction, we first sort
the rows of the matrix according to the number of nonzeros. We then rearrange
the matrix according to increasing order of the number of nonzeros. We also
rearrange the ${\bf x}$ vector and then assign the computation corresponding to the
dense rows to the GPU and the sparse rows to the CPU. The entire ${\bf x}$ vector is
kept at both the CPU and the GPU. This therefore suggests that
when using the work sharing approach, one can divide the computation according to what
computation is more suitable for each of the architectures in the hybrid
platform.

On the CPU, we use the Intel MKL \cite{mkl} library routines and on the GPU we use
the CUSP library routines\footnote{Nvidia CUSP library,
http://code.google.com/p/cusp-library/}. These are known to be offer the best possible
results on each platform.

\subsection{\sgemm}
For the \sgemm$\;$ workload, one can see that computations on various rows of
the input matrices are independent of each other. So, we use a work sharing
model in our hybrid implementation. We use the Intel MKL library \cite{mkl}
for computations on the CPU, and use a row-row method based implementation
developed by us recently in \cite{sgemm} for the GPU computations. The main
implementation difficulty experienced in this workload is to arrive at the
appropriate work shares. The work share would be dictated primarily by the
volume of the output. Since estimating the volume of output is as hard as
actually multiplying the matrices, one has to rely on heuristics to arrive at
the work share. In our implementation, we use the runtime of a CPU alone
implementation and a GPU alone implementation to obtain the work share. 

On the CPU, we use the Intel MKL \cite{mkl} library routines and on the GPU we use
the Row-Row method of matrix multiplication. The row-row method works as
follows \cite{sgemm}. In $C_{m\times n} = A_{m\times p} \cdot
B_{p \times n}$, the $i$th row $C$, denoted $C(i, :)$, is computed as
$\sum_{j \in A(i,:)} A(i, j)\cdot B(j, :)$. This formulation works best
on GPUs for sparse matrices as only those elements that contribute to
the output are  accessed. For more details of the Row-Row method on the
GPU, we refer the reader to \cite{sgemm}.

\subsection{Ray Casting}
In ray casting, rays can perform the computations independently. Therefore,
the problem is very much portable for parallel architectures. Tracing
multiple rays in an SIMD fashion is however challenging because rays access non-contiguous
memory locations, resulting in incoherent and irregular memory accesses.

We notice that there are two main steps in ray casting. The first step is to
find the first triangle that is intersected by each ray. This is then used to
find the first tetrahedron intersected. The second step involves tracing the
ray from the first hit point and traversing the ray through the entire mesh to keep
accumulating the intensity values from the interpolation function. The computation
for this ray finishes  once the ray leaves the mesh. 

In our hybrid implementation, we used a work sharing based solution model
since the computation for each ray can be performed independently. However,
the nature and amount of computation in the above two steps differs
significantly. For this reason, we proceed as follows. We ensure that
computation for each ray finishes the first step before starting computation
for  the  second step for {\em any} ray. The work share across the two steps 
is also varied to reflect the varied nature of the computation across the two
steps. The work share for each step is obtained empirically by studying the
time taken by the CPU and the GPU individually on each of the two steps. We
notice that the optimal work shares across the two steps vary significantly 
depending on the platform.

\subsection{Bilateral Transforms and Convolution}
The workload of bilateral filtering poses interesting challenges.  The
workload has regular memory access patterns, offers good work sharing, 
and is compute bound. Bilateral filtering has the following mathematical
equation on an input image $I$ and $O$ being the output image. 

In our hybrid implementation, we use a combination of task parallelism and
work sharing approach. Notice that bilateral filtering involves computing transcendental 
functions that are very time consuming on the GPU. The number of unique such
function evaluations are however limited by the filter size in case of the
spatial filter, and by the number of different intensity values in the case of
the range filter. The largest filter size of interest is typically $15\times
15$. So, there are only 225 different values that have to be evaluated for the
spatial filter. Given that the intensity values for the images under
consideration were between 0 to 255, the range filter also has only 255 unique
values to the computed.  We perform these computations on the 
the multicore CPU and transfer the results to the GPU. This use of novel task
parallelism in our hybrid implementation proved to be quite beneficial and
serves to illustrate the advantages of a hybrid computing platform. 

Given the values of the transcendental functions in the form of  look-up
tables, applying the filter can be done on parts of the image independently.
So, we use the work sharing approach and divide the input image $I$ into two
parts: $I_{\textrm{CPU}}$ and $I_{\textrm{GPU}}$. The CPU and the GPU
computations apply the filter on the above image parts respectively. The size
of the partitions is arrived at empirically. 

On both the CPU and the GPU, our implementation proceeds by each
thread reading a set of pixels and applying the filter using the look-up
tables. Additionally on the GPU, we make use of shared memory by having each
thread block load the required image from the global memory to the shared
memory.

For the {\em convolution} workload, we observe that the computation is very
regular, compute bound,  and is also highly amenable to data parallel 
operations. As is the standard approach in most GPU implementations and
other modern architectures, we imagine that each thread is computing on a
small portion of the image. To take advantage of the hybrid computing
platform, we divide the computation into two parts according to a certain
threshold. The multicore CPU computes the convolution of a part of the image
$I_{CPU}$ and the GPU computes on the rest of the image, say $I_{GPU}$. The
threshold is chosen as follows. It is observed, (cf. \cite{isca10}), that
the GPU has about a 3x time advantage compared to multicore CPUs. As the
exact model of the CPU and the GPU used in the study of \cite{isca10} is
comparable to that of our platform, we start with the assumption that the
threshold could be around 25\%. We then fine tune the actual threshold by
experimentation. On the CPU, we utilize the availability of the Intel MKL
implementation that is known to be by far the most efficient implementation
of convolution on multicore CPUs. For the GPUs, we use our own custom
implementation. For more details, we refer the reader to \cite{srs12}.

\subsection{Monte Carlo}
Monte Carlo applications typically involve several iterations each using
pseudorandom numbers to estimate the expected value of a random variable of
interest.  We chose the application of photon
migration from \cite{bcsd02}. In the hybrid solution, we
use a hybrid pseudorandom number generator that we developed in
\cite{lspp12}. 
In photon migration, several photons are launched with their
position and direction initialized to either zeros (for some
pencil beam initialized at the origin) or some random numbers.
At every step a photon takes, a fraction of its
weight is absorbed, and then photon packet is scattered. The
new direction and weight of photon are updated. After several
such steps if the remaining weight of a photon is below a
certain threshold, the photon is terminated.

\subsection{List Ranking and Connected Components}
For the list ranking workload, we summarize the approach used in
\cite{hipc11,lspp12}. We start by preprocessing the linked list to reduce 
the size of the list from $n$ nodes to a size of $n/\log n$ using ideas from
fractional independent sets. We then use the algorithm of Hellman and JaJa
\cite{HJ99} to rank the list of remaining elements. Finally, the ranking is
extended to the elements removed from the list during the preprocessing
phase. For the preprocessing, we require that each node in the linked list
choose a bit among $\{0, 1\}$ independently and uniformly at random. For
this, we use the multicore CPU to generate a stream of pseudorandom
numbers. These numbers are then transfered to the GPU so that each node can
access the pseudorandom numbers when it requires. 

In the \CC$\;$ workload, as reported in \cite{hipc11},
we notice that the best possible algorithms for
sequential processing, such as DFS, and PRAM-style parallel algorithms use
fundamentally different techniques. Therefore, we use the following
approach that divides the computation across the multicore CPU and the
GPU. The input  graph $G$ is partitioned into two induced subgraphs,
$G_1 = G[V_1]$ and $G_2 = G[V_2]$ where $V(G) = V_1 \cup V_2$ and $V_1 \cap V_2 =
\Phi$. We use BFS on the CPU cores to find the connected components in the 
graph $G_1$ and the algorithm of Shiloach and Vishkin \cite{sv} on the
GPU to find the connected components in the graph $G_2$.
In the above computation, edges with exactly one end point in either $V_1$
or $V_2$ are not included. We call such edges as {\em cross edges}. 
Hence, in a final step, the connected components of $G_1$ and $G_2$ are
combined using the cross edges. This final step in done on the GPU. The
size of $V_1$, and hence the size of $V_2$, is fixed using an
experimentally obtained threshold.

\subsection{LBM}
The \LBM$\;$ workload is highly parallelizable since each particle can be distributed to
each thread of computation. The Lattice Boltzman model simulates the
propagation and collision processes of a large of a large number of
particles. We perform the simulation over cubic lattices. 
A standard notation in LBM is the $DnQm$ scheme. The parameter $n$ stands
for the dimensions of the cubic lattice, and $m$ stands for the number of
"speeds" studied. In this work, we study the D3Q19 lattice
model where over a 3-dimensional cubic lattice, each particle computes the new
function values based on its present values. 

The computations of various functions are independent of each other. So, 
in our hybrid implementation, we use a task parallel solution approach.
Given the relative speeds of the CPU and the GPU, 
we choose to compute four functions on the CPU and the 
remaining 15 functions are computed on the GPU. Each GPU thread is assigned
the computation with respect to one particle. This can be seen to improve the
data coalescing effects on GPU.

\subsection{Image Dithering and Bundle Adjustment}
For the image dithering workload, we use different strategies to 
perform FSD on  the CPU  and the GPU. On multi-core CPUs, we
formulated a block based approach as this reduces the total threads
required, which is favorable to the multi-core CPUs. For many-core GPUs,
we operate at the pixel level since many-core GPU architecture is suited
for larger number of light weight threads. We notice that  as we use both
the CPU and the GPU in a work sharing model, we need to transfer at most 
three floating point numbers from the CPU to the GPU. We therefore were
able to arrive at an efficient hybrid solution. More details of this
implementation appear in \cite{hipc11pjn}.

For the bundle adjustment workload, we decompose the LM algorithm into multiple
steps, each of which is performed using a kernel on the GPU or a function on
the CPU. Our implementation efficiently schedules the steps on CPU and GPU
to minimize the overall computation time. The concerted work of the CPU and
the GPU is critical to the overall performance gain. 
The implementation that we use here appears in \cite{bundle}.

\section{Results and Discussion}
\label{sec:results}

In this section, we start by describing our evaluation methodology, the
results obtained, and then discuss the results in the context of hybrid
computing.

\begin{table*}[htp]
\begin{center}
\begin{tabular}{|c||c|c|c|c|c|}
\hline
{\bf Workload} & {\bf Dataset} & {\bf Hybrid-High} & {\bf
Hybrid-Low } & {\bf Hybrid High } & {\bf Hybrid Low } \\
 & &  {\bf Gain\% } & {\bf Gain\% } & {\bf Idle Time\% } &   {\bf Idle Time\% } \\
\hline
\hline
\sort & uar &  18.6 & 28.9 & 13.3 & 9.1  \\
\hline  
\hist & uar & 32.3 & 21.8 & 47.7 & 27.9 \\
\hline  
\spmv & \cite{vuduc07} & 15.1 & 48.4 & 1.2  & 1.1 \\
\hline  
\sgemm &  \cite{vuduc07} & 38.9 & 41.87 &  1.76 &    0.41\\
\hline  
\RC & \cite{RMBOF07,LGMBP08} &  23.8 & 39.7 &  10.3 &  8.9 \\
\hline  
\LBM & uar & 15.0 &  11.6 &  27.4 &  23.5  \\
\hline  
\bilat & uar &  12.9 & 7.22 & 1.5 & 1.5  \\
\hline  
\conv & uar &  23.5 &  41.0 & 0.04 & 0.04  \\
\hline   
\MC & uar &   15.7& 16.8 &  6.0 & 4.78  \\
\hline  
\LR & uar &  57.7 &  33.9 & 4.2 & 9.11 \\
\hline   
\CC & \cite{rmat} & 45.16 &   56.4 &  2.85 &  3.76  \\
\hline  
\Dither & uar &  25.5 & 10.5 & 8.9 & 6.42 \\
\hline  
\Bundle & \cite{ndurl} &  88.4 &  78.8 & 77.0 & * \\
\hline  
\end{tabular}
\end{center}
\caption{{\sc Summary of results of our implementations on the Hybrid-High and the
Hybrid-Low platforms. The phrase "uar" in the second row refers to the 
dataset that contains
items drawn uniformly at random appropriate for the workload. A citation in
the second row indicates that we have used the datasets from the work
cited. The performance gain indicated is according to the following metric: time
for \sort$\;$ and \hist, GFlops for \spmv, time for \sgemm, Frames per second
for \RC, time for \LBM, M pixles/sec for \bilat$\;$ and \conv, and time for \MC, \LR,
\CC, \Dither, and \Bundle. For the \Bundle$\;$ workload, the idle time on
Hybrid-Low platform is not available.}}
\label{tab:results}
\end{table*}

\subsection{Evaluation Methodology}
On the platforms described in Section \ref{sec:prelim}, we have used  OpenMP
specification 3.0 \cite{openmp} and CUDA 4.0 \cite{cuda} to implement
our hybrid solutions. Our GPU
programs also use standard optimization techniques such as coalesced
memory access, use of shared memory, minimizing thread divergence, and
the like. The hybrid programs are optimized according to best 
practices for hybrid solutions such as minimizing the overall execution 
time, minimizing the idle time for any device, asynchronous transfer 
of intermediate values between the devices, and the like.  

We are interested in two aspects of hybrid solutions. Firstly, we want
to study the benefits of hybrid computing. We define the {\em gain} of a
hybrid solution as the ratio of the  time taken by the hybrid program to the
minimum time required by a pure GPU or a pure CPU solution. 
Secondly, by nature, hybrid solutions should
minimize the amount of time any of the device is idle. We  define the 
{\em idle time} of a hybrid solution as the total time any device in the hybrid platform
is not used in the computation. This could be due to waiting for results
from other device, or not alloted any part of the computation, or not
alloted enough part of the computation.  A low idle time
indicates better resource efficiency.

\subsection{Results}
\begin{figure*}[htp]
\centering
\subfigure[Sorting]{\includegraphics[scale=0.4]{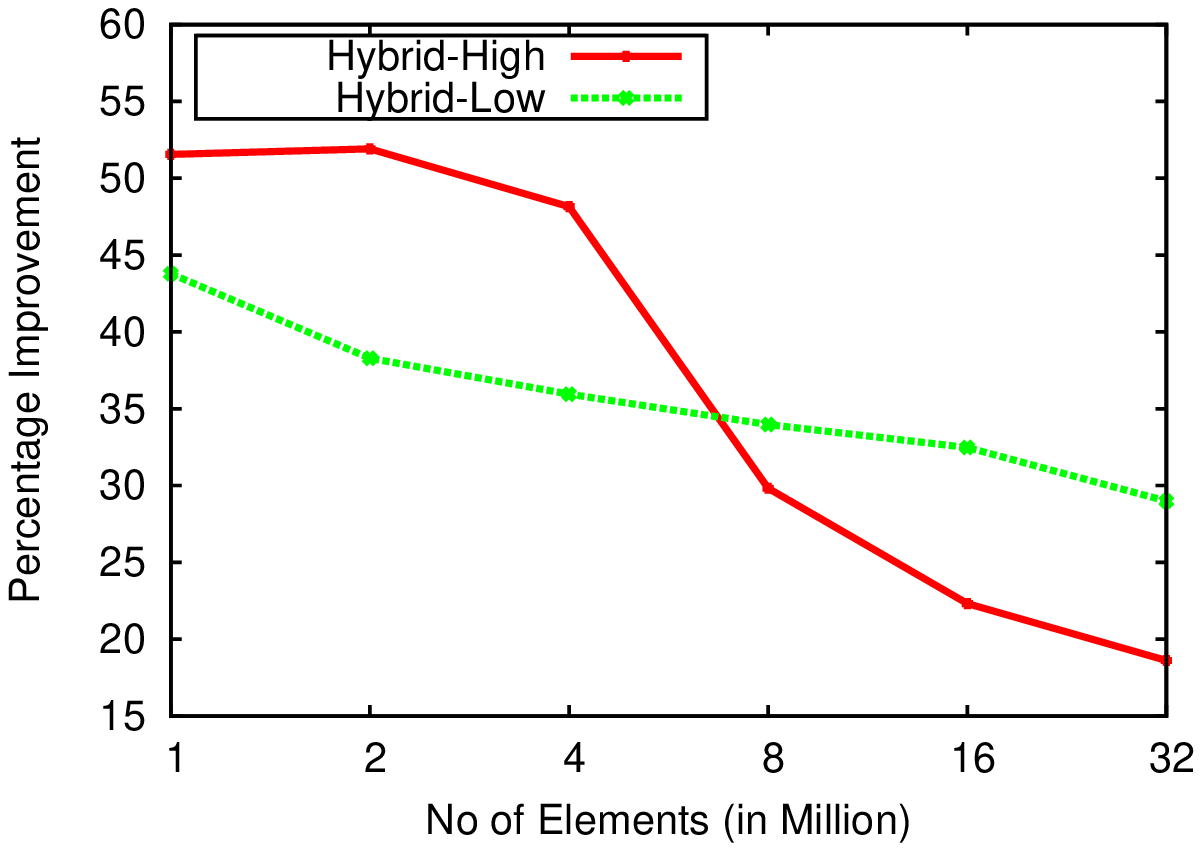}}
\subfigure[Histogram]{\includegraphics[scale=0.4]{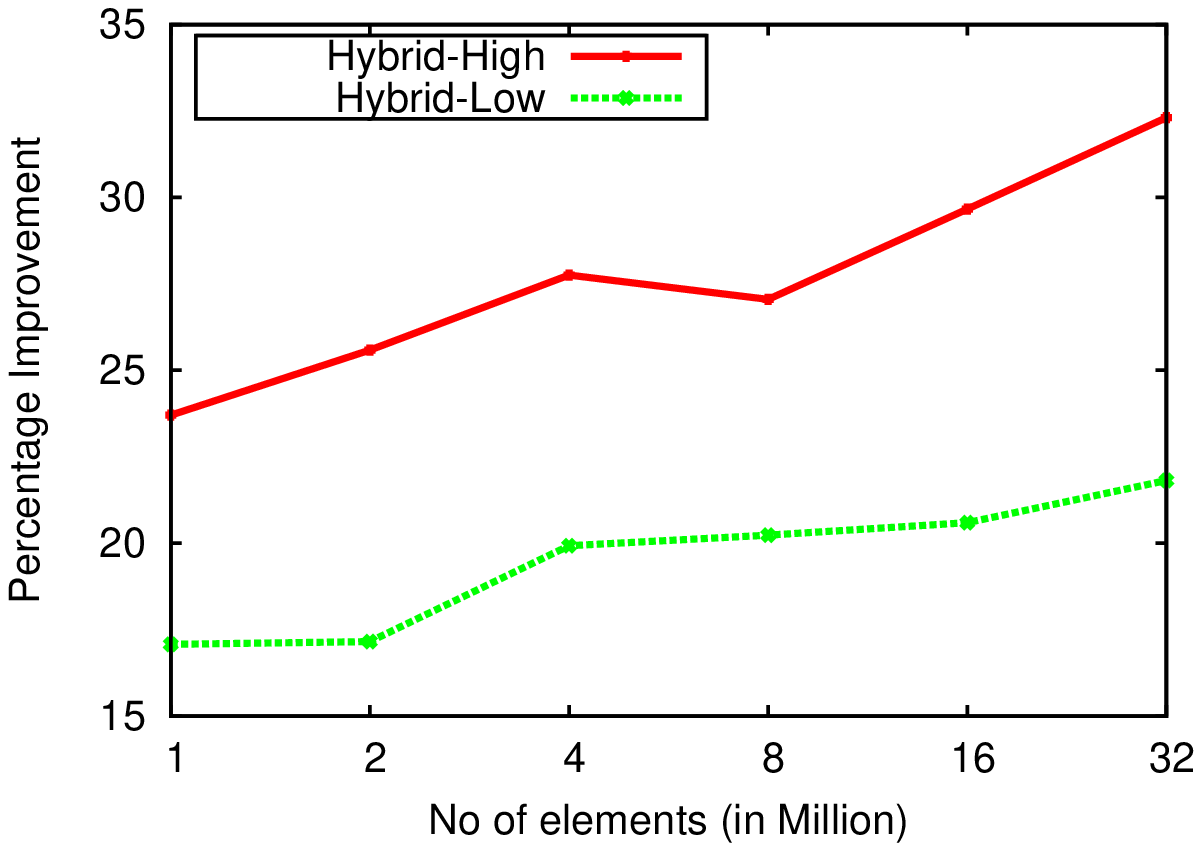}}
\subfigure[\spmv]{\includegraphics[scale=0.4]{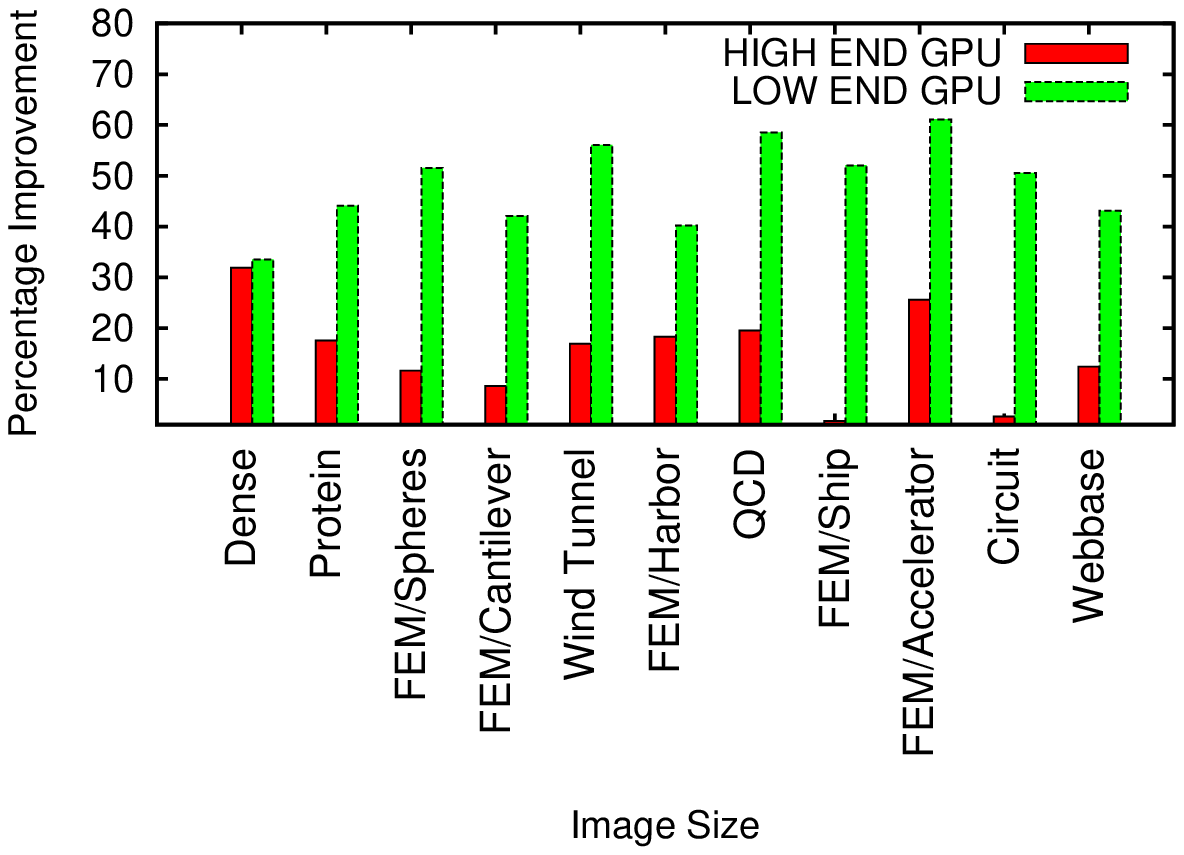}}
\hspace{.1in}
\hfill
\subfigure[\sgemm]{\includegraphics[scale=0.4]{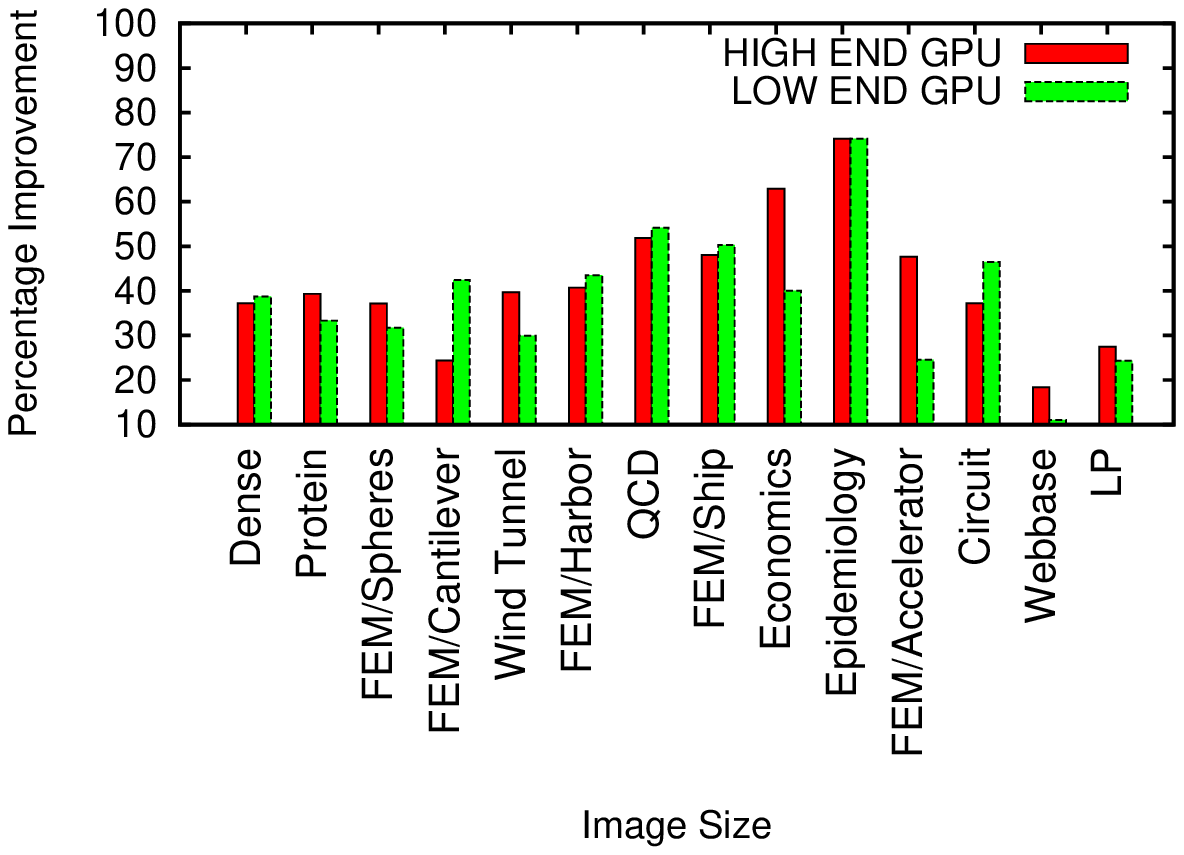}}
\subfigure[Ray casting]{\includegraphics[scale=0.4]{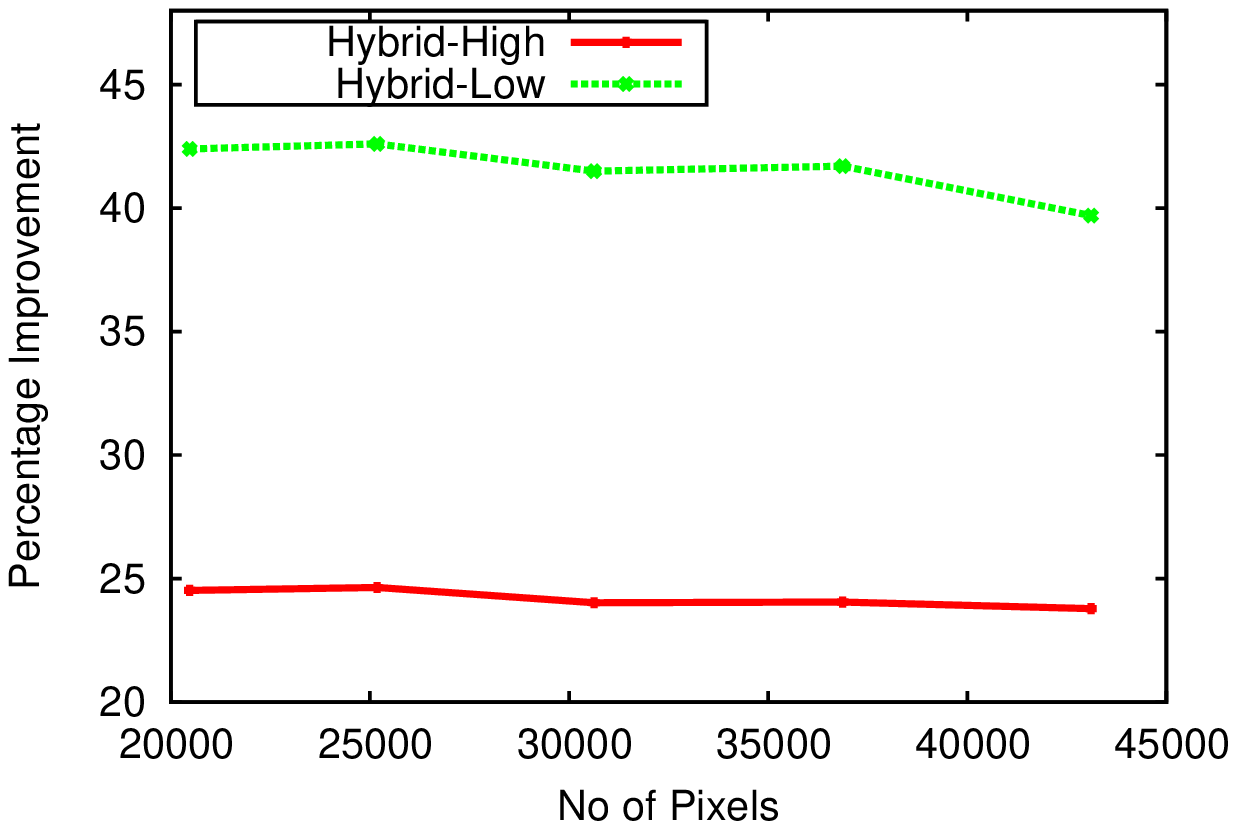}}
\subfigure[Bilateral Filtering]{\includegraphics[scale=0.4]{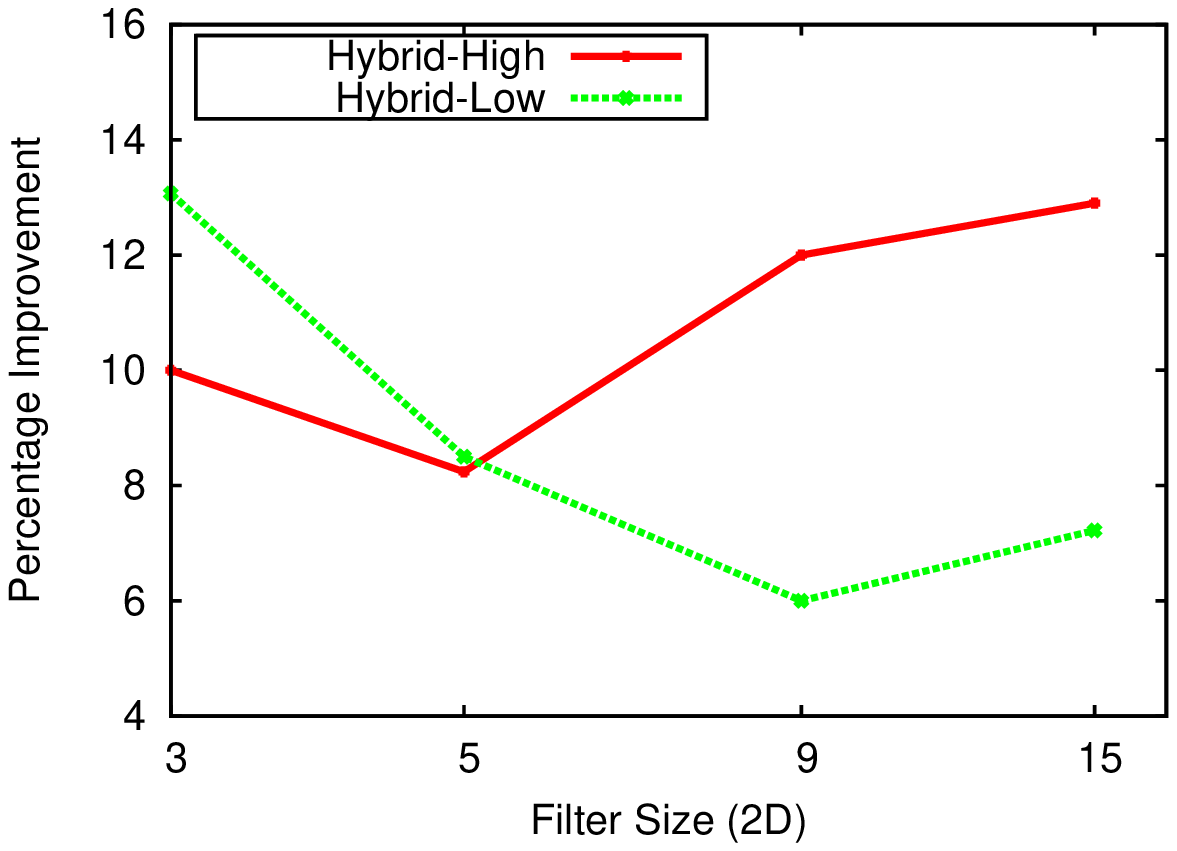}}
\hspace{.1in}
\hfill
\subfigure[Convolution]{\includegraphics[scale=0.4]{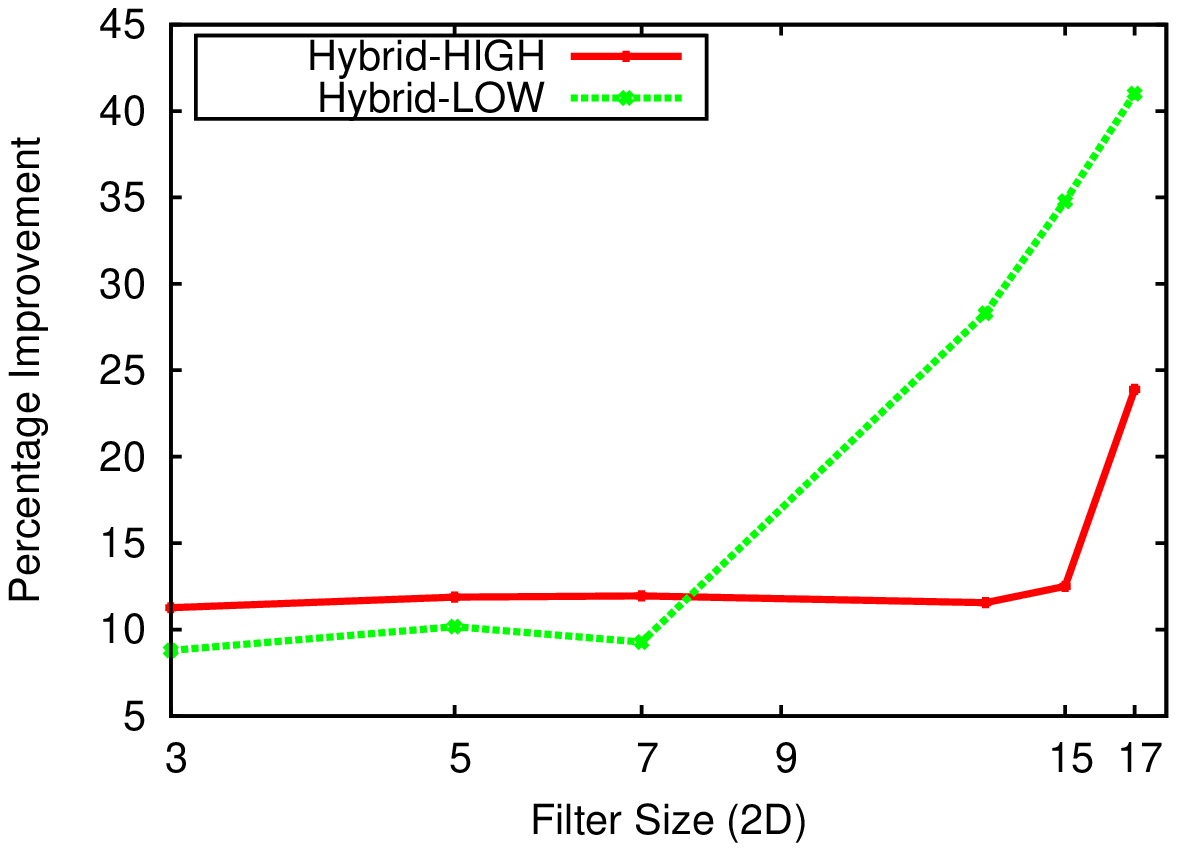}}
\subfigure[Monte Carlo]{\includegraphics[scale=0.4]{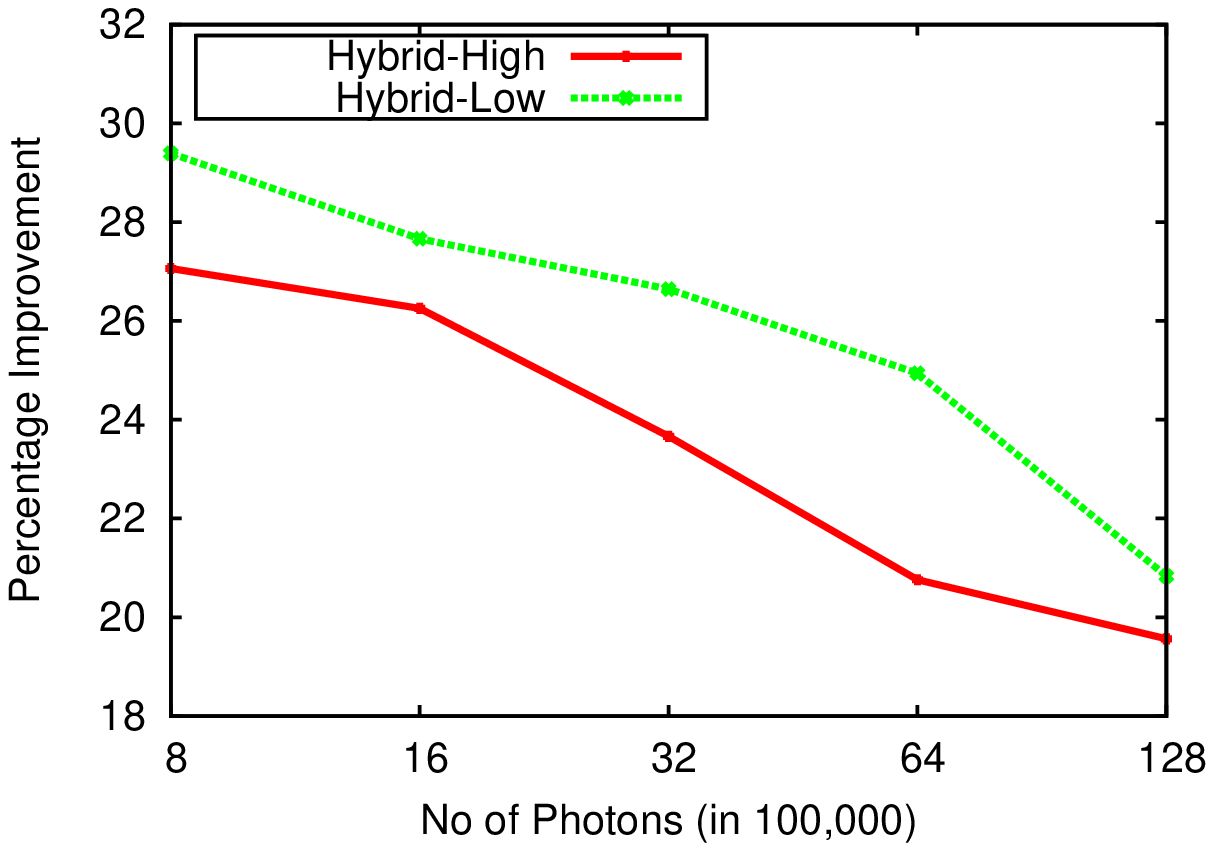}}
\subfigure[List ranking]{\includegraphics[scale=0.4]{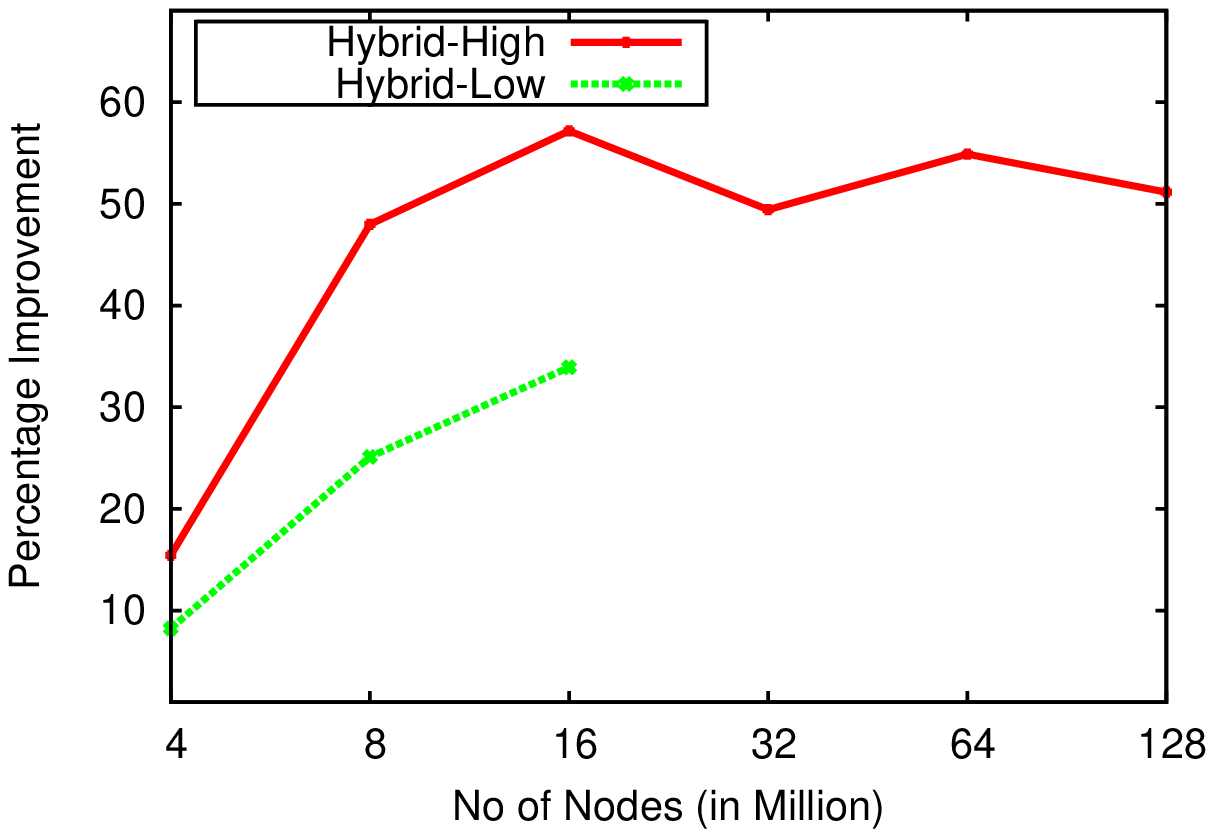}}
\hfill
\subfigure[Connected components]{\includegraphics[scale=0.4]{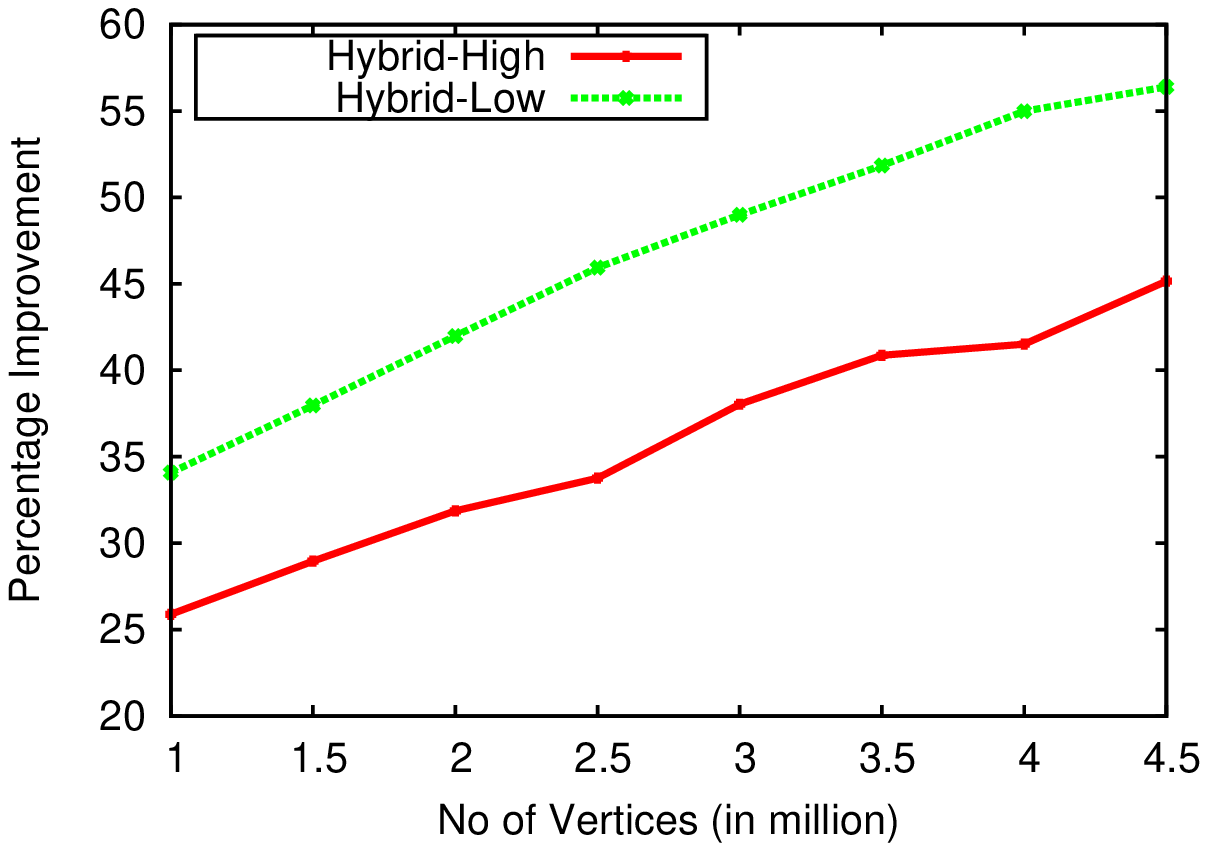}}
\subfigure[Lattice-Boltzman Method]{\includegraphics[scale=0.4]{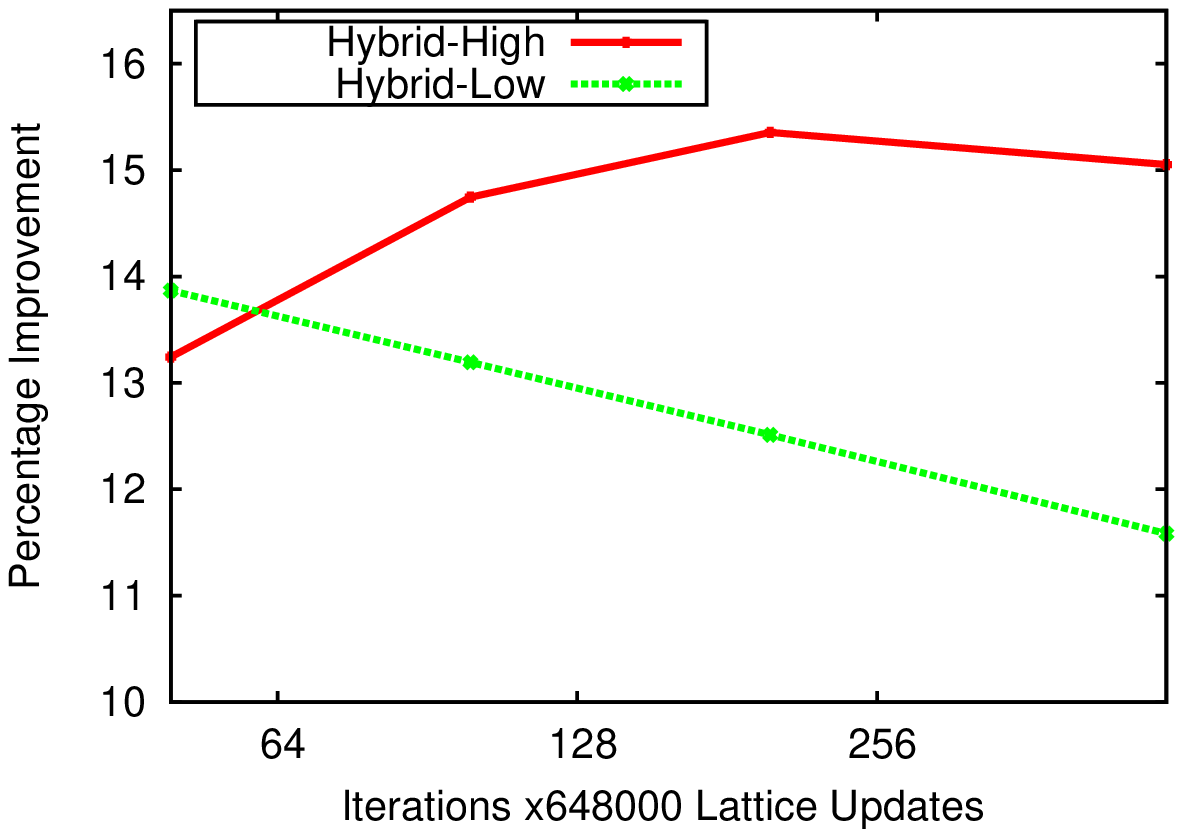}}
\subfigure[Image Dithering]{\includegraphics[scale=0.4]{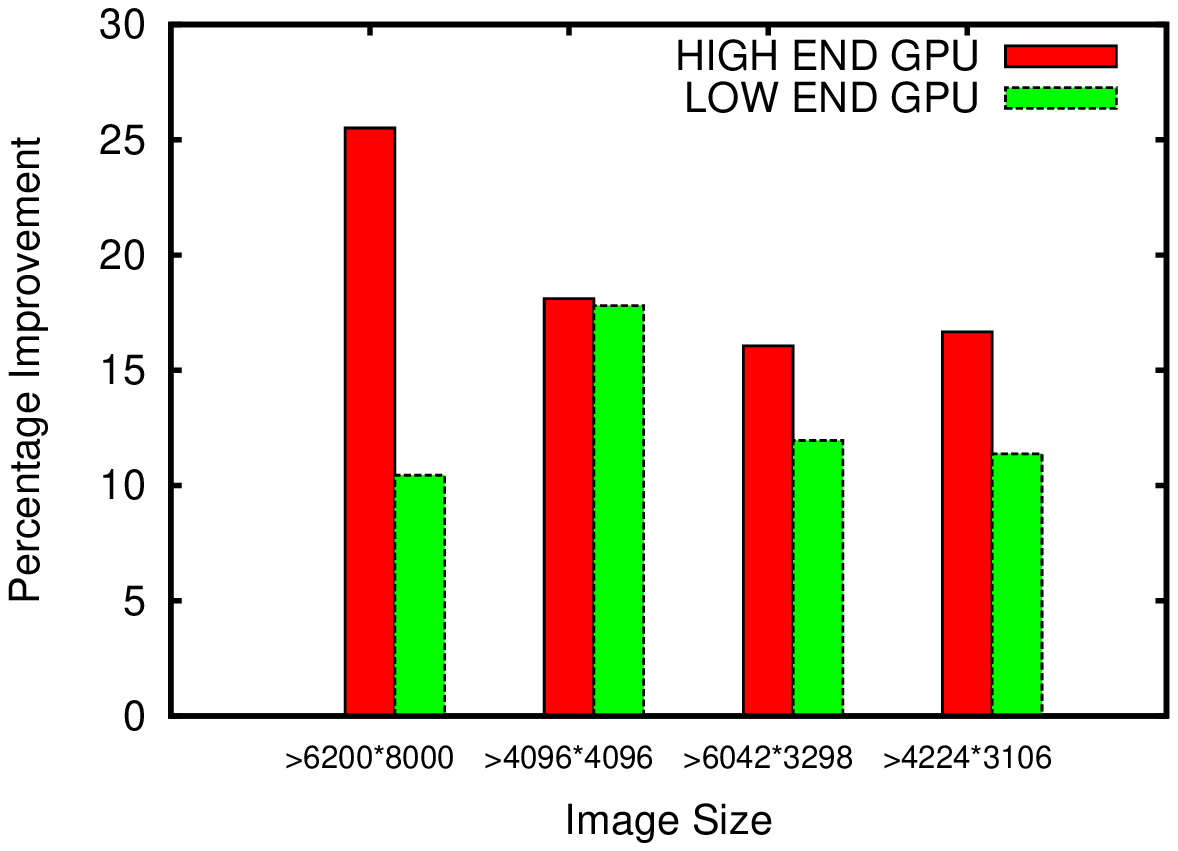}}
\caption{The plots show the performance improvement (in percentage) of
hybrid solutions over a pure GPU solution for the workloads considered
over various input sizes.}
\label{fig:plots}
\end{figure*}

Table \ref{tab:results} summarizes the results of our hybrid solutions.
The second row of Table \ref{tab:results} specifies the dataset used in our
study. The entries in the third and the fourth row are the percentage
improvement of hybrid solutions using the Hybrid-High and the Hybrid-Low
platforms respectively.  The fifth and the sixth row of the Table
shows the idle time of our hybrid solutions on both the platforms
considered. The performance gain and the idle times are for the largest
input sizes for all workload except \spmv, \sgemm. For these two workloads
we  use the average measurement over all the instances in the dataset
considered \cite{vuduc07}.
The values reported in Table \ref{tab:results} and Figure
\ref{fig:plots}[a]--[l] are the average over multiple runs. The results of
Table \ref{tab:results} indicate that our hybrid solutions offer an average
of 30\% improvement on the Hybrid-High platform, and an average of 34\% on
the Hybrid-Low platform. Remarkably, the Hybrid-Low platform whose
configuration is likely to match commonly used desktop configurations also
offers good incentives for hybrid computing. 

Figure \ref{fig:plots}[a]--[l] show the performance of our hybrid
implementations on various inputs from the datasets mentioned in the
second row of Table \ref{tab:results}. 
The plots in Figure \ref{fig:plots} show that our hybrid implementations
scale well over increasing input sizes. In most cases, our maximum input 
size is limited only by the available memory on the GPU in the hybrid
platform. 

On most workloads, our results on the Hybrid-High and Hybrid-Low
platforms suggest that hybrid computing has scope and advantage.  Our
workloads also have applications in common settings such as graphics and
image manipulation, data processing, and the like. Some of these
operations are invoked internally by regular users of computers such as
gamers. The input sizes that we used in evaluating the Hybrid-Low
platform are also close to the typical usage in most cases. 

\subsection{Analysis of the Results}
In this section, we analyse the results shown in Section
\ref{sec:results} in the context of our evaluation methodology.

\subsubsection{Performance of Hybrid Solutions}

Some of our workloads such as  \conv, \bilat$\;$ are
very amenable to the GPU style of computation. On such workloads, as can
be noticed from Table \ref{tab:results}, the hybrid advantage on the
Hybrid-High platform is rather modest. This is accentuated by the fact that
the GPU in the Hybrid-High platform has a peak throughput that is 10 times
that of the CPU in the Hybrid-High platform when considering single
precision operations.  On the Hybrid-Low
platform however, as the ratio of the peak throughput of the GPU and the
CPU is smaller, hybrid computing can be seen to offer a decent advantage
even on regular workloads.

On the other hand, about half our workloads have irregular memory access
patterns that are
known to be very difficult for GPUs to handle. Examples of such workloads
include \LR, \sgemm, \CC, and the like. 
On such workloads, as can be observed from Table \ref{tab:results}, hybrid
computing on the Hybrid-High platform offers better than 40\% advantage on 
workloads such as \LR$\;$ even while the GPU peak throughput is 
about ten times that of the peak CPU throughput.
This is possible by using novel task mapping techniques
that assign the right task to the right processor. This suggests that as
GPUs suffer on workloads with highly irregular memory access patterns,
one should think of utilizing the power of hybrid computing. 

For the workloads that are common to the workloads considered in
\cite{isca10}, our GPU alone results are either better or comparable than
those reported in \cite{isca10}. Since the CPU used in \cite{isca10} is
different from the CPUs we used in our hybrid computing platforms, it is not
possible to compare the CPU alone performance. 

\subsubsection{Idle Time}
Table \ref{tab:results} also shows the idle time of our workloads on
both the platforms. For workloads that use a work sharing parallel
approach, it can be observed that the idle time is quite small. 
This is due to the fact that at the right threshold of work
distribution, the CPU and the GPU  take near identical times. 

For workloads using a task parallel solution approach, such as \LBM,
and \Bundle, it is possible that the computation time is not matched
between the CPU and the GPU. In the case of \LBM$\;$ and \Bundle, further
fine-tuning of the task assignment is also not possible. In the
case of Bundle adjustment workload, there is no equivalent Pure-GPU code as
the hybrid code is a direct extension of the available CPU code.
 Some tasks are
not amenable to a further sub-division which means that computation on those
tasks would always result in a imbalance on the CPU and the GPU runtime.
In such cases, the idle time tends to be high.

\subsection{Discussion}

In this section, we try to highlight some of the lessons that were
learnt during our study in hybrid computing. These can offer some
insights into how future heterogeneous architectures at the commodity 
scale and also at the higher end can be designed.

\subsubsection{Communication Cost}
In most of the hybrid computing solutions, it is required that the
devices transfer intermediate results or other such data related to the
progress of the computation. For instance, in the sorting workload, in
our hybrid implementation,  as the GPU further splits bins which have
more than a pre-selected number of elements, the CPU is sorting the bins
with fewer elements. To enable this, we send the starting and ending
indices of the bins with fewer elements that the CPU can sort.

Ideally, one likes to hide this communication with computation. However,
at present, computing on CPU-GPU hybrid platforms is difficult as the
communication bandwidth between the CPU and the GPU is via the PCI
Express link. On the Hybrid-High
platform, the peak bandwidth offered is about 6 GB/s. This limitation
means that hybrid solutions have to think of novel ways to minimize the
amount of communication, hide communication latencies with other
computations, and possibly avoid communication.  These may also limit the
nature of techniques that can be used in hybrid solution design. 

In future, therefore, one has to conceive hybrid architectures with a
more tighter coupling between the devices so that communication costs can
be minimized. Emerging models such as the Intel MIC and the AMD Fusion 
may offer some hope in this direction and deserve a careful future study.

\subsubsection{The Right Solution Methodology}
In Section \ref{sec:intro}, we have identified two broad solution
methodologies that hybrid algorithms use, namely task parallelism and
work sharing. As we use these two approaches for the 13 workloads
presented in this paper, we discuss which approach may be suitable for a  given
problem. 

The work sharing solution methodology involves dividing the work between
the CPU and the GPU so that both take roughly the same computation time. 
This solution methodology is useful when the computation on a part of the
input is almost independent of the computation on the other part of the
input. For instance, in the \conv$\;$ workload, the computation on each pixel
is dependent only on the values of the neighboring pixels in the input
image. This property allows computation on one sub-image to be treated
independent of the rest of the image. Similar observations apply to the
\spmv$\;$ workload. 

The task parallel approach is useful when the computation can be seen as
a set of tasks and their dependencies. It may be useful to also
represent the tasks and their dependencies as a task graph naturally.
Further, the tasks should be such that there exists tasks that are more
efficient on a particular architecture. For instance, in the \LR$\;$
workload, we identify tasks such as generating pseudo-random numbers and
computing a fractional independent set (FIS) as two tasks. However,
dependent tasks executing on different devices implies that the results
of one task have to be necessarily communicated to the other task. The
communication time has to be taken into account when mapping the tasks
to the devices.

\subsubsection{Identifying the Right Threshold in Work Sharing} 
The work sharing approach to hybrid computing
suggests that the CPU and the GPU in the hybrid platform split the
overall work in some ratio. This solution methodology is used in
workloads such as \sort, \hist, \MC, \sgemm, \bilat, and \conv. 

\begin{figure}
\centering
\epsfig{file=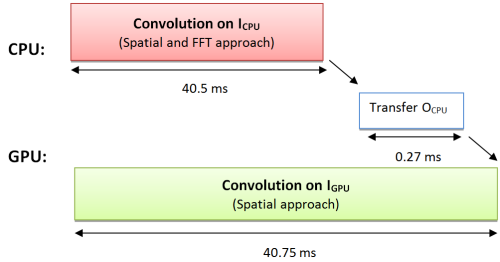, scale=0.25}
\caption{Figure showing the CPU and GPU overlapped computation on the \conv$\;$ hybrid
solution on a 3600$\times$ 3600 image and a 15$\times$15 filter.}
\label{fig:conv}
\end{figure}

In this case, one can see that the work
distribution should ideally be according to the ratio of the 
processing times on the CPU and the GPU in the platform. For instance, if the
GPU alone runtime is $T_{GPU}$ and the CPU alone runtime is
$T_{CPU}$, then the hybrid solution should split work as
$\frac{T_{GPU}}{T_{GPU}+T_{CPU}}$ percentage on the CPU and the
remaining percentage on the GPU. The calculation indicates an ideal
scenario where all intermediate communication is hidden by useful
compute, and no post-processing of the partial results obtained by the
CPU and the GPU is required. An example is shown in Figure \ref{fig:conv}
where the input image is split in a ratio of 18\%.
One can therefore use the above calculation to identify a good work
distribution to start with and then 
adjust it experimentally after also taking into account the communication
times and the post-processing involved. 

\subsubsection{Identifying and Mapping Work Units in the Task Parallel
Approach}
Some of our hybrid solutions use the technique of task parallelism. In this
technique, we identify work units, or tasks, and their inter-dependence in terms
of their precedences. These tasks are then mapped onto the best possible
device according to the architectural suitability. We discuss two issues
in this context that affect the performance of hybrid solutions.

Firstly, it is not easy in general to identify the right tasks, as computing is
often traditionally understood in a sequential step-by-step manner. Even in
parallel computing, the intention in general is to speed up each step of the
computation using the available processors. Only recently are other
methodologies for parallel computing such as using domain specific languages
\cite{dsl1,dsl2} are gaining attention. While these languages
alleviate the job of writing efficient parallel programs, they can still be
constrained by a traditional step-by-step approach of problem solving.

Identifying the tasks and their dependencies requires a careful
reinterpretation of the computation involved. For instance, in the \bilat$\;$
workload, we noticed that GPUs are not amenable to computing transcendental
functions. These were therefore executed on the CPU. Further, it is also
noticed that there are really very few transcendental function evaluation
that are required for a given image. (These are based on the maximum
difference between the pixel intensities). Therefore, we precompute these
values, and transfer these values from the CPU to the GPU. While we may be
precomputing more values than needed in an actual input, the benefits of
this model stem from the fact that recomputing transcendental functions is
rather expensive on any architecture. 

\begin{figure}
\centering
\epsfig{file=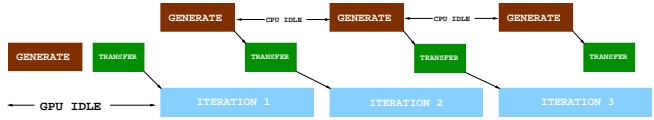, scale=0.15}
\caption{Figure showing the assignment of tasks during the \LR$\;$ hybrid
solution on a list of size 128 M elements.}
\label{fig:LRwork}
\end{figure}

The \LR$\;$ workload offers similar insights. Our implementation of
list ranking in a hybrid setting \cite{hipc11} has a preprocessing phase
thar requires a large quantity of random numbers.  These random numbers can be
generated on the CPU and transfered to the GPU. In our implementation, we
generate the random numbers on the CPU and the GPU uses the 
random numbers thus supplied. This is seen to save a lot of processing time
in the hybrid setting. Figure \ref{fig:LRwork} shows the
task assignment used in our hybrid implementation \LR.

Secondly, it is not easy to identify the right task for the right
processor. At present, our arguments are based on intuitive reasoning backed
by experimental evidence. In future, we would like to study
formal mechanism to arrive at an appropriate and near-optimal task mapping.
In fact, arriving at an optimal assignment can be easily seen to be an
NP-complete problem and hence one should consider near-optimal
assignments.

\ignore {
\subsubsection{Programming Models}

Hybrid computing requires a significant reinterpretation of programs to
achieve performance improvements.  Current programming models offer
little support for writing hybrid programs. In all the workloads that we
implemented for this study, we used OpenMP \cite{openmp} and the CUDA
programming model for programming mutlicore CPUs and NVIDIA GPUs
respectively. In light of this, it is important that the gains in
efficiency offset the programming effort involved in writing hybrid
programs. This also suggests that one should consider developing
appropriate programming models that ease the programming effort on
hybrid computing platforms.
}

\subsubsection{Lessons for other Hybrid Computing Platforms}
Heterogeneity in architectures is a common phenomenon in
recent times. Therefore, hybrid computing is poised to
play a great role so as to improve resource efficiency. It is hence
important to understand how applications can treat
the heterogeneity as an advantage. In this section,
we extrapolate the lessons learnt from this paper for other hybrid
computing platforms and architectural recommendations for the same.

Our experience with the Hybrid-Low platform suggests that the computing
devices in a hybrid platform should have similar peak performance
capabilities. However, heterogeneity helps by allowing the right task to
be executed on the right device. The current trend in equipping
architectures with special purpose accelerators such as MMX units, CRC
units, encrypt/decrypt units, therefore allows for more task parallel
hybrid solutions. 
Having dedicated accelerators can also make processor design less
complex, and also allows for simpler frequency scaling thereby improving 
power efficiency.

\section{Related Work}
\label{sec:related}
There has been considerable interest in GPU computing in recent years. Some
of the notable works include scan \cite{scan}, \spmv \cite{BG09},
sorting \cite{sanders10}, and the like. Other modern
architectures  that have been studied recently 
include the IBM Cell and the multi-core machines. Bader et al.
\cite{bader} have studied list ranking on the Cell architecture
and show that by running multiple threads at each SPU, list ranking
using the Hellman-JaJa algorithm can be done efficiently. Other notable
works on the Cell architecture include \cite{cellBFS,vuduc07}. Williams et
al. \cite{vuduc07} have studied the {\tt spmv} kernel on
various multi-core architectures including those from Intel, Sun, and
AMD. 
Since most of the above cited works do not involve hybrid computing, 
we do not intend to cite all such works in this paper and refer the 
reader to other natural sources. 

A recent work that motivated this paper is the work of Lee et al.
\cite{isca10}.
In \cite{isca10}, Lee et al. argue that GPU computing can offer on average
only a 3x performance advantage over a multicore CPU on a range of 14
workloads deemed important for throughput oriented applications. Some of
our workloads overlap theirs ~\cite{isca10}. Their
paper also generated a wide amount of debate on the applicability and limitations
of GPU computing. Our view however is that it is not a question of whether
GPUs can outperform CPUs or vice-versa, but rather what can be achieved
when GPUs and CPUs join forces in a viable hybrid computing platform. 
Further, for the workloads that are included also in \cite{isca10}, we
provided our own GPU and CPU implementations. In workloads such as \bilat, 
we use novel ideas such as precomputing the transcendentals on the
GPU for a pure GPU implementation  that improve the performance beyond what is
reported in \cite{isca10}.

Hybrid computing is gaining popularity across application areas such as
dense linear algebra kernels \cite{BDT08,dongarra2009,AICCSA11}, maximum flows
\cite{hong09},  graph BFS
\cite{kunle11} and the like. The aim
of this paper is to however evaluate the promise and the potential of
hybrid computing by considering a rich set of diverse workloads. 
Further, in some of these works, (cf.
\cite{kunle11,hong09,WJ10}), while both the CPU
and the GPU are used in the computation, one of the devices is idle and
while the other is performing computation. In contrast, we seek
solutions where both the devices are simultaneously involved in the
computation.

There have been recent works that propose benchmark suites for GPU
computing. Popular amongst them are the Rodinia \cite{rodinia} and SHOC
\cite{shoc}. Some of our workloads such as sorting, \sgemm$\;$
are part of the SHOC Level one benchmark suite.
Subsets of the workloads considered in our paper appear in other
benchmarking efforts related to parallel computing. The Berkeley report
\cite{landscape} lists dwarfs as computational patterns that have wide
application. Workloads such as \sort, \hist, {\tt spmv}, {\tt
sgemm},  are part of Berkeley dwarfs. This
serves to illustrate the wide acceptance of our chosen workloads.

\section{Conclusions}
\label{sec:concl}
In this paper, we have evaluated the case for hybrid computing
by considering workloads from diverse application areas and two different
hybrid platforms. We also
experimented with two hybrid platforms and analyzed their suitability 
for hybrid computing. 
Our study opens the way
for evaluation on other challenges with respect to hybrid computing such as 
power efficiency, benchmark suites, and performance models
for hybrid computing (see \cite{hipc09,kim09}).

\bibliographystyle{IEEEtranS} 

\bibliography{sc} 

\end{document}